\documentclass[draft]{agujournal2019}
\usepackage{url} 
\usepackage{lineno}
\usepackage[inline]{trackchanges} 
\usepackage{soul}
\usepackage{verbatim}
\draftfalse

\journalname{JGR: Space Physics}

\begin{document}

%
%


\title{Three-Dimensional Magnetic Reconnection Spreading in Current Sheets of Non-Uniform Thickness}

%
%




\authors{Milton Arencibia\affil{1}, P. A. Cassak\affil{1}, M. A. Shay\affil{2}, Jiong Qiu\affil{3}, Steven M.~Petrinec\affil{4}, Haoming Liang\affil{5}}

\affiliation{1}{Department of Physics and Astronomy and Center for KINETIC Plasma Physics, West Virginia University, Morgantown, WV 26506, USA}
\affiliation{2}{Department of Physics and Astronomy, University of Delaware, Newark, Delaware 19716, USA}
\affiliation{3}{Department of Physics, Montana State University, Bozeman MT, 59717, USA}
\affiliation{4}{Lockheed Martin Advanced Technology Center, Palo Alto, CA 94304 USA}
\affiliation{5}{Center for Space Plasma and Aeronomic Research (CSPAR), University of Alabama in Huntsville, Huntsville, AL 35805, USA}




\correspondingauthor{Milton Arencibia}{milton.arencibia@gmail.com}




\begin{keypoints}
\item We derive a theory of three-dimensional spreading of collisionless anti-parallel reconnection in current sheets with non-uniform thickness.
\item Spreading from a thinner to a thicker current sheet occurs slower than local electron and Alfv\'en speeds, a key prediction of the theory.
\item We apply the theory to reconnection spreading in Earth's magnetotail and discuss potential implications for solar flare ribbons.
\end{keypoints}

%
%

%
%


\begin{abstract}
   Magnetic reconnection in naturally occurring and laboratory settings often begins locally and elongates, or spreads, in the direction perpendicular to the plane of reconnection.  Previous work has largely focused on current sheets with a uniform thickness, for which the predicted spreading speed for anti-parallel reconnection is the local speed of the current carriers. We derive a scaling theory of three-dimensional (3D) spreading of collisionless anti-parallel reconnection in a current sheet with its thickness varying in the out-of-plane direction, both for spreading from a thinner to thicker region and a thicker to thinner region. We derive an expression for calculating the time it takes for spreading to occur for a current sheet with a given profile of its thickness.  A key result is that when reconnection spreads from a thinner to a thicker region, the spreading speed in the thicker region is slower than both the Alfv\'en speed and the speed of the local current carriers by a factor of the ratio of thin to thick current sheet thicknesses. This is important because magnetospheric and solar observations have previously measured the spreading speed to be slower than previously predicted, so the present mechanism might explain this feature. We confirm the theory via a parametric study using 3D two-fluid numerical simulations. We use the prediction to calculate the time scale for reconnection spreading in Earth's magnetotail during geomagnetic activity.
   The results are also potentially important for understanding reconnection spreading in solar flares and the dayside magnetopause of Earth and other planets. 
\end{abstract}

\section*{Plain Language Summary}
Magnetic reconnection is fundamental process in plasmas that converts magnetic energy into kinetic and thermal energy and is known to mediate eruptive solar flares and geomagnetic substorms that create the northern lights.  The x-line where magnetic reconnection occurs can elongate or spread over time in the direction normal to the plane of reconnection, and this trait has been observed in the laboratory, Earth's magnetosphere, and is thought to be related to the elongation of chromospheric ribbons during solar flares. This study presents a scaling theory of the 3D spreading of anti-parallel magnetic reconnection in current sheets with thickness varying in the out-of-plane direction. A key result is that when reconnection spreads from a thinner to a thicker region, the spreading speed in the thicker region is slower than expected. This is important because magnetospheric and solar observations have observed slower spreading speeds than previously predicted, so the present mechanism might explain this feature. We confirm the theory with 3D numerical simulations and use the prediction to calculate the time scale for reconnection spreading in Earth's magnetotail during geomagnetic activity.

%
%

%


%
%
%
%

\section{Introduction}
\label{sec-intro}

The abrupt release of magnetic energy in substorms in Earth's magnetosphere and flares in the solar corona are key features of the dynamics of these systems and have an important impact on Earth's technological infrastructure. In both processes, magnetic reconnection is the driver of the rapid energy conversion \cite{McPherron73,Priest00} via a change in magnetic field connectivity \cite{Dungey53,Vasyliunas75}. Observations have revealed that  reconnecting x-lines (the collection of points where the magnetic field connectivity changes) often start in a localized region of space, and then elongate or spread in time, orthogonal to the reconnection plane in two-ribbon solar flares \cite{Isobe02,Qiu09,Qiu10,Tian15,Graham15,Qiu17} and prominence eruptions \cite{Tripathi06}, at Earth's magnetopause \cite{Zhou17,Zou18, walsh18}, in Earth's magnetotail \cite{McPherron73,Nagai82,Nagai13,Hietala14}, and in laboratory reconnection experiments \cite{Katz10,Egedal11,Dorfman13}. Reconnection starting locally and spreading is also thought to happen in the solar wind where x-lines hundreds of Earth radii in extent have been observed \cite{Phan06,Gosling07c,Shepherd17}. 

Most of the previous theoretical and numerical work on the spreading of reconnection has addressed quasi-2D anti-parallel reconnection in uniform current sheets with an initial half-thickness $w_0$ comparable to the ion inertial scale $d_i = c/\omega_{pi}$, where $c$ is the speed of light in vacuum and $\omega_{pi}$ is the ion plasma frequency.
The consensus is that reconnection spreads orthogonal to the reconnection plane with the velocity of the current carriers \cite{huba02,Huba03,Shay03,Karimabadi04,Lapenta06,Shepherd12,Nakamura12,Mayer13,Jain13,Jain17,arencibia21}. This directionality of the spreading is consistent with observations of reconnection during substorms, which spread in the dawnward direction \cite{McPherron73,Nagai82,Nagai13}.  While the ions carry most of the current in the quiet plasma sheet, the electrons carry the current when the plasma sheet thins down when reconnection takes place \cite{Jain21}, so the direction of the spreading is consistent with the direction of the current carriers. 

However, reconnecting current sheets in naturally occurring physical systems such as the solar corona and the dayside magnetopause and magnetotail of Earth and other planets are unlikely to have a thickness that is uniform in the out-of-plane direction before reconnection onsets and spreads. For example, {\it in situ} observations of the near-Earth magnetotail plasma sheet during quiet times show the half-thickness varies continuously in the dawn-dusk direction from a minimum of $< 3 \ R_E$ at midnight in magnetic local time up to $\sim 8 \ R_E$ at the flanks, and thins down to $\sim 0.1-0.4 \ R_E$ at midnight and $\sim 1 \ R_E$ at the flanks at the end of a substorm growth phase, prior to reconnection onset \cite{fairfield79,fairfield80,voigt84,sergeev90,kaymaz94,tsyganenko98}, where $R_E$ denotes the radius of Earth. Thin current sheets where reconnection is more likely to occur are more prevalent on the dusk-side of the magnetotail \cite{Rong11,Rogers22}. Interestingly, however, reconnection is suppressed within 10 $d_i$ of the duskward edge of the region undergoing reconnection \cite{Liu19}, so magnetotail reconnection need not be strongest at the thinnest part of the current sheet. At Earth's dayside magnetopause, the current sheet is thinnest near the nose and gets thicker towards the flanks \cite{Haaland14}, so reconnection spreads in a non-uniform current sheet. Moreover, in situ observations suggest that magnetosheath high speed jets can trigger dayside reconnection where the magnetopause current sheet is as thick as 60-70 $d_i$ \cite{Hietala18}, as corroborated by numerical simualtions \cite{Ng21}, so dayside reconnection also need not begin at the thinnest part of the current sheet.
In solar flares, intermittency of the sequential brightening of ribbons has been interpreted as evidence of the nonuniformity in the out-of-plane direction of the flare current sheet \cite{naus22}. It has been seen in global magnetospheric simulations that reconnection spreading slows as reconnection spreads from a thinner to a thicker current sheet \cite{walsh18}. While there have been numerical studies of a current sheet of non-uniform thickness that was extremely thick outside the reconnection region so that the x-line remained spatially confined \cite{Liu19,Huang19}, we are unaware of any studies that predict the spreading speed of reconnection in current sheets of a non-uniform thickness.

We present a scaling theory of the spreading of collisionless anti-parallel reconnection in current sheets of non-uniform thickness. We include predictions for spreading from a thinner to thicker current sheet and for spreading from a thicker to thinner current sheet. Reconnection may start at its thinnest part, but magnetotail observations suggest that this need not be the case, so both limits are potentially physically relevant. For reconnection that spreads from a thinner into a thicker part of a current sheet, a key result is that the spreading speed in the thicker region is slower than the spreading speed based on current knowledge for a uniform sheet of equivalent local thickness, due to a reduction in the initial effective reconnecting field \cite{Shay04}. This provides a mechanism for reconnection spreading that is sub-Alfv\'enic as well as slower than the local current carriers in the macroscopic current sheet. This result is important because observations of dayside reconnection \cite{Zou17} and two-ribbon solar flares \cite{Qiu17} suggest that the spreading speed is slower than expected from the existing theory. We confirm our prediction with a suite of 3D two-fluid numerical simulations. We use our prediction for the spreading speed to calculate the time it takes reconnection to spread a particular distance. We apply our results to reconnection in Earth's magnetotail, and motivate potential observational signatures of spreading in current sheets of non-uniform thickness in solar flares.

The layout of this paper is as follows. In Sec.~\ref{sec-theory}, we present a theory of 3D reconnection spreading in current sheets of non-uniform thickness. In Sec.~\ref{sec-simulation}, we discuss our numerical simulation setup. In Sec.~\ref{sec-results}, we discuss the results of our simulations. In Sec.~\ref{sec-app}, we apply our results to reconnection in the near-Earth magnetotail and two ribbon solar flares, and offer conclusions in Sec.~\ref{sec-conc}.

\section{Theory}
\label{sec-theory}

We define a coordinate system in which $z$ is the out-of-plane direction coincident with the direction of the initial current density $J_z$, $x$ is the direction of the equilibrium reversing magnetic field, and $y$ completes a right-handed coordinate system.  We use a reference frame where the electrons fully carry the initial current for simplicity and treat collisionless reconnection.  We assume the reconnecting magnetic field $B_x$ asymptotes to a magnitude of $B_0$ at all values of $z$ for simplicity. 

\begin{figure} 
\includegraphics[width=\textwidth]{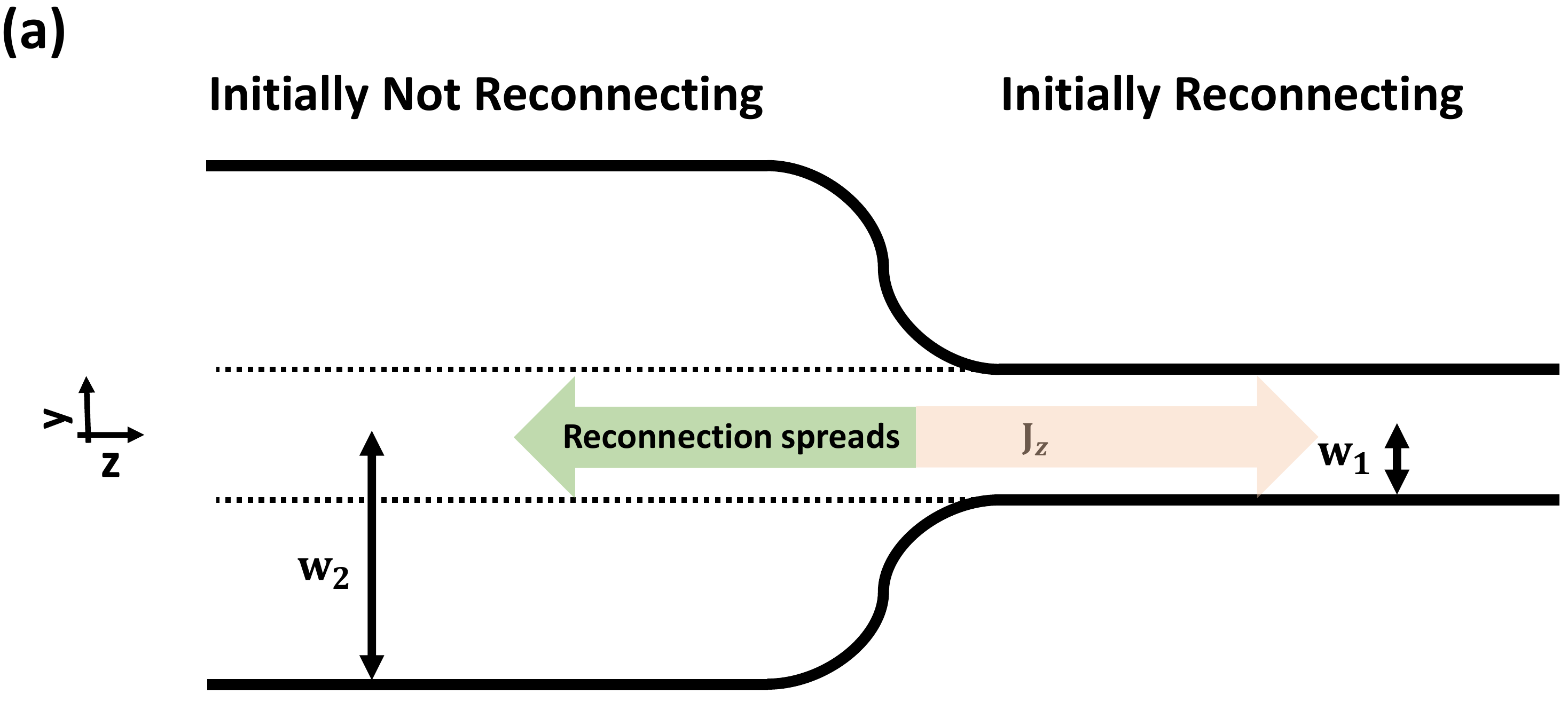}
\vskip 1cm
\includegraphics[width=\textwidth]{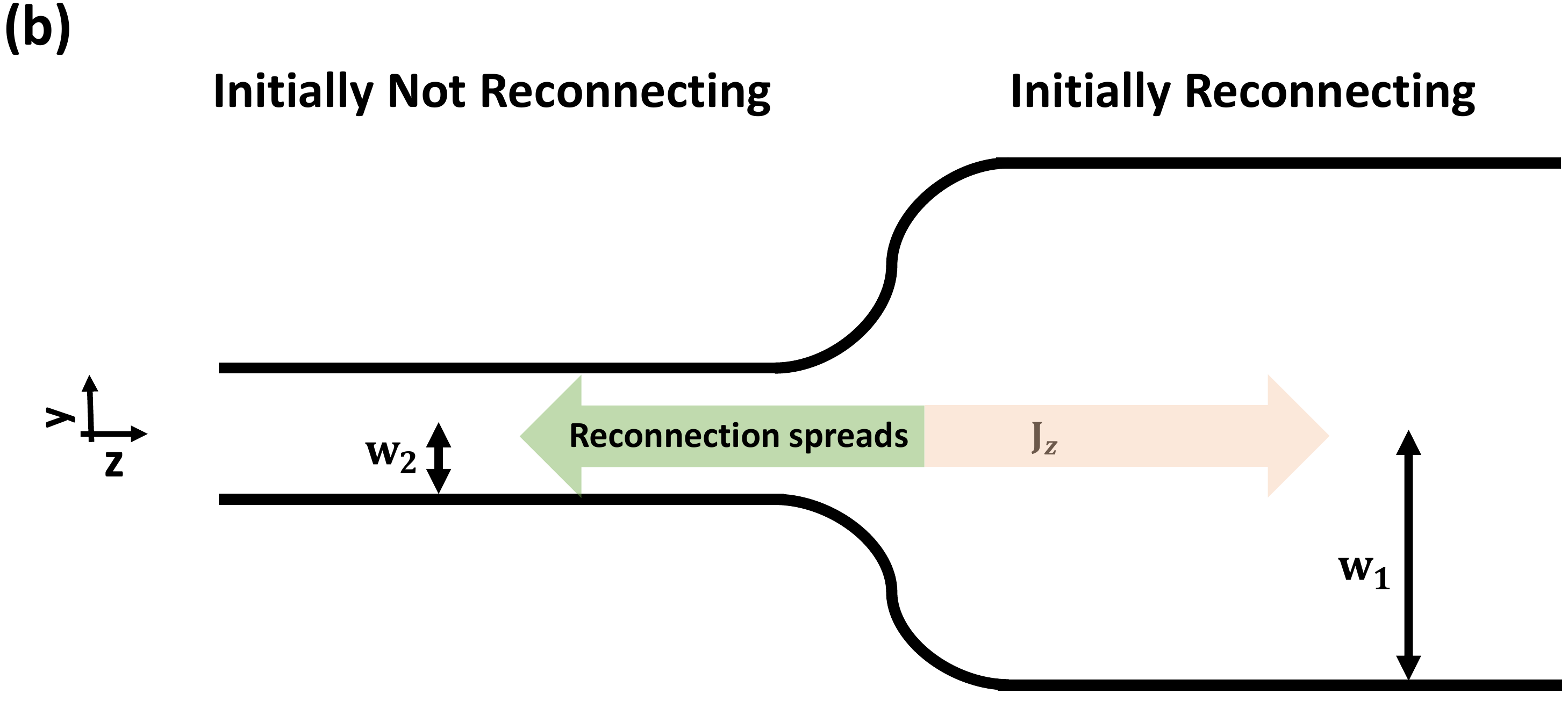}
  \caption{Sketch of the $yz$ plane for a current sheet with non-uniform thickness in the out-of-plane direction in which reconnection spreads from a region of local half-thickness $w_1$ into a region of half-thickness $w_2$, propagated by the electrons carrying the current. Panel (a) is for $w_1 < w_2$ and panel (b) is for $w_1 > w_2$. The dotted lines in panel (a) denote that the reconnected fields that convect into the thicker region essentially remain collimated to a half-thickness $w_1$ from the thinner region, leading to embedded reconnection because the upstream magnetic field is weaker there.}
  \label{fig-sketch}
\end{figure}

The current sheet has a half-thickness $w(z)$ in the $y$ direction that varies in the out-of-plane direction. Representative sketches of the current sheet profile in the $yz$ plane are shown in Fig.~\ref{fig-sketch}.  The solid black lines represent the edge of the equilibrium current layer, and the green arrow denotes the direction that reconnection spreads due to the electron current carriers.  Panel (a) depicts a current sheet with a monotonically increasing current sheet half-thickness, for which reconnection spreads from a thinner to thicker region of the current sheet, while panel (b) depicts a current sheet with a monotonically decreasing current sheet half-thickness, for which spreading is from a thicker to thinner region.  We define the current sheet half-thickness where reconnection starts as $w_1$.  In the simulations we use to test the theory, the current sheet half-thickness asymptotes to $w_2$. We assume the half-thickness $w(z)$ varies slowly as a function of $z$, and we will quantify this condition in what follows. We first introduce general aspects of the derivation of the spreading speed.  Then, we separately calculate the spreading speed as a function of $w(z)$ for monotonically increasing and decreasing thickness profiles.

\subsection{General Considerations of Reconnection Spreading in Current Sheets of Non-Uniform Thickness}
\label{subsec-gentheory}

We begin with a review of the analysis of the spreading speed for anti-parallel reconnection in a current sheet of uniform thickness in \citeA{arencibia21}. Spreading occurs because the reconnected ($y$) component of the magnetic field propagates in the direction of the current carriers, triggering the sequential onset of reconnection and causing the x-line to grow in length. This is governed by Faraday's law, given in cgs units as
\begin{equation}
    \frac{\partial B_y}{\partial t} \simeq -c \frac{\partial E_x}{\partial z}, \label{eq:faraday}
\end{equation}
where ${\bf B}$ is the magnetic field, ${\bf E}$ is the electric field, and the variation of $E_z$ in the $x$ direction is assumed small. In a small interval of time $\Delta t$, $B_y$ propagates a distance $\Delta z$ in the $z$ direction, and the spreading speed is defined as $v_s = \Delta z / \Delta t$. It is estimated from a scaling analysis of equation~(\ref{eq:faraday}), giving
\begin{equation}
    v_s \sim -c \frac{\Delta E_x}{\Delta B_y}. \label{eq:vsuniform1}
\end{equation}
For anti-parallel reconnection, it was argued that the main contributor to $E_x$ is the Hall electric field $-J_z B_y / nec$ \cite{arencibia21}, where $n$ is the upstream density and $e$ is the elementary charge, so that
\begin{equation}
    v_s \sim c \frac{\Delta (J_z B_y / nec)}{\Delta B_y}. \label{eq:vsuniform2}
\end{equation}
For a current sheet of uniform thickness, $J_z$ and $n$ are independent of $z$, so equation~(\ref{eq:vsuniform2}) becomes
 \begin{equation}
    v_s \sim \frac{J_z}{ne}. \label{eq:vsuniform3}
\end{equation}
This result provided a first-principles scaling prediction of the previously known result that spreading of anti-parallel reconnection in a current sheet of uniform thickness occurs at the speed of the current carriers \cite{huba02,Huba03,Shay03,Jain13}.  Since $J_z \sim c B_0 / 4 \pi w$, equation~(\ref{eq:vsuniform3}) gives 
\begin{equation}
v_s \sim \frac{c B_0}{4 \pi n e w} \sim \frac{c_{A0} d_i}{w}, \label{eq:vsuniform6}
\end{equation}
where $d_i = c/\omega_{pi} = (m_ic^2 / 4\pi n e^2)^{1/2}$ is the ion inertial scale, $c_{A0} = B_{0} / (4 \pi n m_i)^{1/2}$ is the Alfv\'en speed based on $B_0$, and $m_i$ is the ion mass.

We now show how to generalize this theory for spreading in a current sheet of non-uniform thickness, where $v_s$ is expected to be a function of $z$.  Equations~(\ref{eq:faraday}) - (\ref{eq:vsuniform2}) are unchanged, but when $w(z)$ is non-uniform, $J_z(z)$ is no longer uniform.  Continuing to treat $n$ as uniform for simplicity, using the chain rule in equation~(\ref{eq:vsuniform2}) gives
\begin{equation}
    v_s(z) \sim \frac{1}{ne} \left[ J_z(z) + B_y(z) \frac{\Delta J_z(z)}{\Delta B_y(z)} \right], \label{eq:vsuniform4}
\end{equation}
where we have made the $z$ dependence explicit.  Since $B_y \simeq 0$ in the non-reconnecting region, $B_y$ is on the same order of magnitude as $\Delta B_y$, so the second term in the brackets scales like $\Delta J_z$ while the first scales like $J_z$.  If $w(z)$ varies rapidly, the second term would need to be retained and may even dominate.  However, since we are assuming that $w(z)$ varies slowly, we argue that $\Delta J_z \ll J_z$, and the second term can be neglected.  In this limit, the spreading speed is
\begin{equation}
    v_s(z) \sim \frac{J_z(z)}{ne} \simeq \frac{c_A(z) d_i}{w(z)}. \label{eq:vswkb}
\end{equation}
where we use Amp\`ere's law to write $J_z(z) \simeq c B_x(z) / 4 \pi w(z)$ and define the Alfv\'en speed as a function of $z$ as $c_A(z) = B_x(z) / (4 \pi nm_i)^{1/2}$, where the reconnecting magnetic field $B_x$ can depend on $z$. We argue in what follows that $B_x(z)$ depends on whether the current sheet half-thickness is increasing or decreasing.

\subsection{Spreading From a Thinner to a Thicker Current Sheet}
\label{subsec-thintothick}

We first consider the system sketched in Fig.~\ref{fig-sketch}(a), with reconnection beginning in a region of uniform half-thickness $w_1$ that spreads  monotonically into a thicker region. We argue
that the reconnected magnetic field $B_y$ is collimated at a half thickness near $w_1$ as it spreads into the thicker current sheet. Reconnection initiates in a plane of given $z$ when $B_y$ appears at that $z$. In the next increment in time $\Delta t$, $B_y$ convects a small distance $\Delta z$. Since the equilibrium current sheet thickness increases with $z$, the half-thickness expands from $w(z)$ to $w(z+\Delta z)$. 
The perturbing $B_y$ is expected to also broaden as it goes from $z$ to $z + \Delta z$. If the vertical inflow speed $v_y$ due to reconnection exceeds the speed of the broadening of $B_y$ due to the increase of the half-thickness of the current sheet, then $B_y$ remains collimated as it convects to $z + \Delta z$. Data supporting this statement will be given in Sec.~\ref{sec-results}.

We now develop a quantitative condition for $B_y$ to remain collimated.
Between the thin and thick parts of the current sheet, the reconnecting magnetic field $B_x$ has a gradient in the $z$ direction that is associated with a vertical equilibrium current $J_{y,eq} = (c/4\pi) \partial B_x/ \partial z$.
The associated vertical velocity $v_{ey,eq}$ due to the equilibrium flow which serves to broaden $B_y$ scales as
\begin{equation}
    v_{ey,eq} \sim -\frac{J_{y,eq}}{ne} \sim -\frac{c}{4\pi n e}\frac{\partial B_x}{\partial z}. \label{eq:veyeq}
\end{equation}
We estimate this speed at $y = w(z)$.  Suppose the $B_x$ profile is written as $B_x(y,z) = B_0 \tilde{B}_x[y/w(z)]$, where $\tilde{B}_x$ is a dimensionless function capturing the spatial structure of $B_x$. Using the chain rule, $\partial B_x / \partial z = B_0 \tilde{B}_x^\prime [\partial (y/w)/\partial w] (dw/dz)$, where $\tilde{B}_x^\prime$ is the derivative of $\tilde{B}_x^\prime$ with respect to its argument. Evaluating this at $y = w$ gives $\partial (y/w)/\partial w = -1/w$, so we get $\partial B_x[y=w(z),z] / \partial z = -B_0 \tilde{B}_x^\prime(1) (1/w)(dw/dz) = -B_0 \tilde{B}_x^\prime(1) / w_z$, where $w_z = [d \ln(w) / dz]^{-1}$ is the scale size over which the half-thickness of the current sheet changes. Then, the scaling of equation~(\ref{eq:veyeq}) gives
\begin{equation}
    v_{ey,eq} \sim  \frac{c_{A0} d_i \tilde{B}_x^\prime(1)}{w_z}.
\end{equation}

Letting the inflow speed associated with the reconnection be $v_{in}$, we find that $B_y$ remains collimated if $v_{in} > v_{ey,eq}$, {\it i.e.,} 
\begin{equation}
    w_z > \frac{c_{A0}}{v_{in}} \tilde{B}_x^\prime(1) d_i. \label{eq:deltazcondition}
\end{equation}
$v_{in}/c_{A0}$ is a proxy for the reconnection rate, which we expect to be on the order of 0.1. Since $\tilde{B}_x^\prime(1)$ is typically of order 1, we find that the condition that $B_y$ remain collimated as it enters a thicker current sheet is $w_z$ is at least around 10 $d_i$.  For most physical systems of interest, this is a small scale compared to the size of upstream structures, so it is likely this condition is satisfied. If such small scale structure did occur, it would be prone to kinetic instabilities that smooth out sharp gradients.  Consequently, we expect $B_y$ to remain collimated at the thinner scale $w_1$ as it spreads into regions with a thicker current sheet.  This collimation is sketched as the dotted lines in Fig.~\ref{fig-sketch}(a).

This result implies the magnetic field outside the region of half-thickness $y > w_1$ is not significantly perturbed by $B_{y}$ and does not initially participate in the reconnection in the thicker region,  provided the time scale for spreading is shorter than the time scale for the current sheet to collapse due to reconnection. Consequently, the effective upstream magnetic field that controls the driving of the reconnection process is weaker than the asymptotic magnetic field $B_0$.  Reconnection for which only a thinner sublayer participates in the reconnection process has previously been referred to as ``embedded'' \cite{Shay04,Cassak09b}.

We can estimate the spreading speed semi-empirically. We hypothesize that the effect of embedding is that $B_x(z)$ is lower than $B_0$ in equation~(\ref{eq:vswkb}) for the spreading speed. We estimate the reconnecting magnetic field $B_{x}(z)$ that initially participates in reconnection
by assuming $B_x$ varies approximately linearly in $y$ within the current layer \cite{Shay04}, so that the reconnecting magnetic field in a current sheet of thickness $w(z)$ is
\begin{equation}
B_x(z) \sim B_{0} \frac{w_1}{w(z)}.  \label{eq:buppred}
\end{equation}
The subsequent spreading speed from equation~(\ref{eq:vswkb}) using $c_A(z) = B_x(z) / (4\pi m_i n)^{1/2}$ is
\begin{equation}
    v_s(z) = \frac{c_{A0} d_i w_1}{[w(z)]^2}. \label{eq:caz}
\end{equation}
Therefore, the predicted spreading speed $v_{s}(z)$ at a position $z$ is slower than the spreading speed for a current sheet of equivalent uniform half-thickness $w(z)$, given by $c_A d_i/w(z)$, by a factor of $w_1 / w(z)$.  This is a key prediction of this theory and a departure from previous knowledge of reconnection spreading in current sheets of uniform thickness. It shows the spreading speed fundamentally depends not just on the local current sheet half-thickness $w(z)$, but there is also a ``memory'' effect of the current sheet from where its half-thickness was $w_1$.

\subsection{Spreading From a Thicker to a Thinner Current Sheet}

If reconnection spreads from a thick region into a thinner one as sketched in Fig.~\ref{fig-sketch}(b), the incoming reconnected magnetic field $B_y$ perturbs the entire thickness of the thinner region. The full thickness of the thinner current layer participates in reconnection from the beginning, and thus the relevant upstream magnetic field is the asymptotic magnetic field $B_0$.  This implies that $c_A(z) = c_{A0}$, a constant, in equation~(\ref{eq:vswkb}), so the spreading speed prediction is
\begin{equation}
v_s(z) \simeq \frac{c_{A0} d_i}{w(z)}.
\label{nonuniform-speed-gradual-thick-thin}
\end{equation}
This implies that reconnection in this scenario spreads in the thinner region at a speed given by the local current carrier speed. Thus, in contrast to spreading from a thinner to thicker current sheet, spreading from a thicker to thinner current sheet has no memory effect.

\subsection{Time Scale For Spreading a Prescribed Distance}
\label{subsec-timescale}

Since the speed is a function of position for spreading in non-uniform current sheets, it is challenging to test the spreading speed prediction numerically, experimentally, or observationally by direct measurement. Thus, we also provide a prediction for the time it takes for spreading to occur over some region, which is likely to be easier to measure.  From elementary mechanics, the time $\tau$ it takes to spread from position $z_1$ to $z_2$ is
\begin{equation}
\tau = \int_{z_1}^{z_2} \frac{dz}{v_{s}(z)}, \label{eq-tau-general}
\end{equation}
where the appropriate form of $v_s(z)$ needs to be used for thinner-to-thicker or thicker-to-thinner current sheet thickness profiles.

While equation~(\ref{eq-tau-general}) is expected to be valid for any gradually changing thickness profile $w(z)$, we exemplify the procedure by assuming a half-thickness profile $w(z)$ of the power law form
\begin{equation}
w(z) = w_1 + (w_2 - w_1) \left(\frac{z}{\Delta z}\right)^\alpha, \label{eq-tailprofile}
\end{equation}
where $\Delta z = z_2 - z_1$, $z_1 = 0$, and $\alpha$ is a dimensionless parameter that can be chosen for a particular model current sheet. Here, $w(z_1) = w_1$ and $w(z_2) = w_2$.  We first consider spreading from a thinner to thicker current sheet.  Using equation~(\ref{eq-tailprofile}) in equation~(\ref{eq:caz}), the integral in equation~(\ref{eq-tau-general}) straight-forwardly gives
\begin{equation}
\tau = 
\frac{w_1\Delta z}{c_Ad_i} \left[ 1 + 2\frac{w_2/w_1 -1}{\alpha+1} + \frac{(w_2/w_1-1)^2}{2\alpha+1}  \right].
\label{eq-tau}
\end{equation}
To interpret this result, we note that the prefactor is the transit time for reconnection spreading in a uniform current sheet of half-thickness $w_1$ over a distance $\Delta z$.  Therefore, the $\alpha$- and $w_2$-dependent terms in the brackets represent a geometric factor which describes the increase in the spreading time due to the current sheet becoming thicker.

Similarly, for reconnection spreading in a current sheet that decreases in half-thickness gradually from $w_1$ to $w_2$ with a profile according to equation~(\ref{eq-tailprofile}), the local spreading speed is instead given by 
equation~(\ref{nonuniform-speed-gradual-thick-thin}). Then, the integral in equation~(\ref{eq-tau-general}), 
after simplifying, gives
\begin{equation}
\tau = \frac{w_1\Delta z}{c_Ad_i} \left( 1 + \frac{w_2/w_1 - 1}{\alpha + 1} \right).
\label{eq-tau2}
\end{equation}

\section{Simulation Setup}
\label{sec-simulation}

The simulation study is carried out using the two-fluid code F3D \cite{Shay04}, which updates the continuity, momentum, induction, and pressure equations, and includes the Hall and electron inertia terms in the generalized Ohm's law to account for separate electron and ion dynamics below the ion inertial scale. Time is stepped forward using the trapezoidal leapfrog algorithm \cite{Guzdar93} and spatial derivatives are fourth order finite differences.  Lengths are normalized to the ion inertial scale $d_{i0} = (m_ic^2/4\pi n_0 e^2)^{1/2}$, time is normalized to the inverse ion cyclotron frequency $\Omega_{ci0}^{-1}=m_ic/eB_0$, velocities to the Alfv\'en speed $c_{A0}=B_0/\sqrt{4\pi m_in_0}$, electric fields to $c_{A0} B_0 / c$, current densities to $c B_0 / 4 \pi d_{i0}$, and temperatures to $m_i c_{A0}^2 / k_B$, where $B_0$ is the initial asymptotic strength of the reversing magnetic field, $n_0$ is the initial upstream density, and $k_B$ is Boltzmann's constant.

For this study, we employ an identical simulation setup as our anti-parallel reconnection simulations in an earlier study \cite{arencibia21} with the exception of a non-uniform current sheet thickness profile. We use a computational domain with dimensions $L_x \times L_y \times L_z = 102.4 \times 51.2 \times 256.0$ with triply periodic boundary conditions.  The grid scale is $ \Delta x \times \Delta y \times \Delta z = 0.05 \times 0.05 \times 1.0$. The time step is $0.02$ and the ion-to-electron mass ratio is $m_i/m_e=25$ for all simulations in this study.

\begin{figure} 
\center \includegraphics[width=\textwidth]{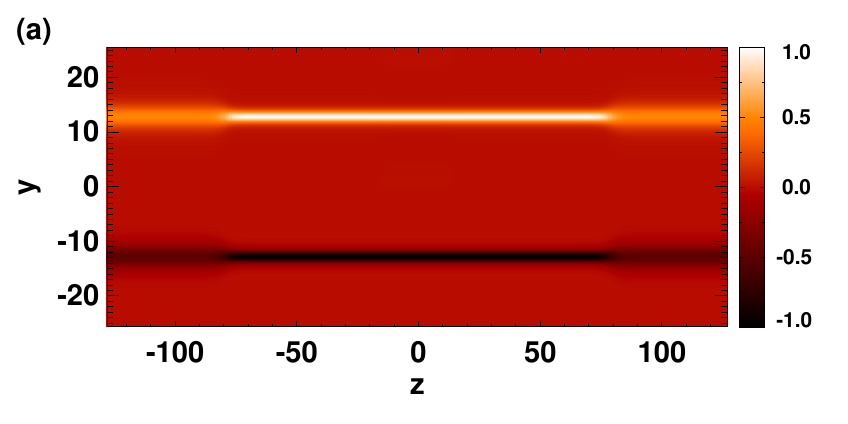}
\center \includegraphics[width=\textwidth]{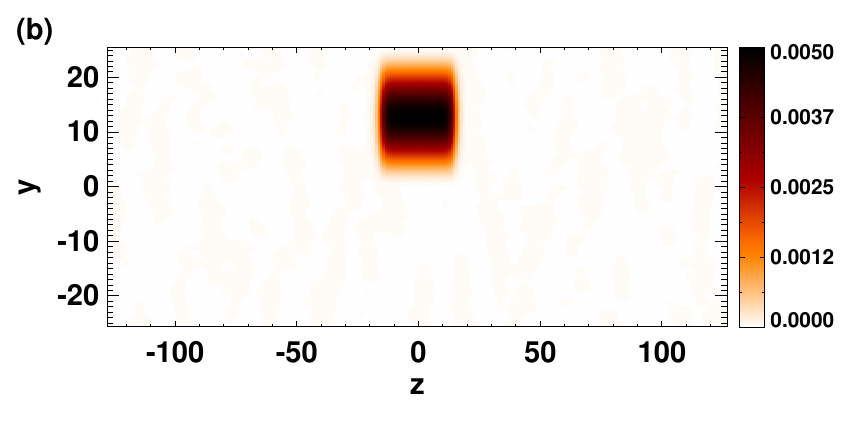}
  \caption{Representative initial conditions for a simulation with a current sheet with thickness that varies in the out-of-plane direction. The plots give a cut in the $yz$ plane at $x = -L_x/2$ of (a) the initial current density $J_z$ with a thickness profile given by equation~(\ref{eq-wofz}) with $w_1 = 1$ and $w_2 = 2$, and (b) the $y$-component of the magnetic perturbation $B_{1y}$. 
  \label{fig-initconds}}
\end{figure}

For our initial conditions, we employ two oppositely directed current sheets with the $x$-component of the initial magnetic field given by $B_{0x}=\tanh[(y + L_y/4)/w_0(z)] - \tanh[(y-L_y/4)/w_0(z)] - 1$, so that the two current sheets are centered at $y = \pm y_{cs} = \pm L_y/4$, which permits the use of periodic boundary conditions. Here, $w_0(z)$ is the initial current sheet half-thickness profile, which varies in the out-of-plane direction between two specified values $w_1$ and $w_2$, given by
\begin{equation}
w_{0}(z) = \frac{w_1 + w_2}{2} + \frac{w_1 - w_2}{2} \left[\tanh \left(\frac{z + L_{0}}{w_z}\right) - \tanh \left(\frac{z - L_{0}}{w_z}\right) - 1\right], \label{eq-wofz}
\end{equation}
where $z = 0$ is the center of the computational domain, $L_0 = 80$ is the half-length in the out-of-plane direction of the region of half-thickness $w_1$, and $w_z = 4$ (unless otherwise stated) is the gradient scale length over which the half-thickness changes from $w_1$ to $w_2$.  The full extent of the region over which the current sheet changes from a half-thickness of $w_1$ to $w_2$ is $\Delta z = 2 w_z$, which for all simulations in this study is large enough to satisfy equation~(\ref{eq:deltazcondition}).  We carry out two suites of simulations, one holding $w_1 = 1.0$ fixed and varying $w_2 = 1.25, 1.5, 1.75, 2$ (all thinner to thicker), and another holding $w_2 = 2$ fixed and varying $w_1 = 0.75, 1.5, 1.75, 1.9, 2.25, 2.5$ (a combination of thinner to thicker and thicker to thinner). We also carry out one simulation with $w_1 = 2.0$ and $w_2 = 1.5$ (thicker to thinner) and an additional two simulations with uniform half-thicknesses $w_0 = 1.0$ and 2.0. Thicker initial current sheets are desirable but, because they take longer to evolve, are significantly more computationally expensive. Fig.~\ref{fig-initconds}(a) shows initial conditions for the out-of-plane current $J_z$ in a cut in the $yz$ plane at $x=-L_x/2$, showing distinct regions of different half-thicknesses $w_1 = 1$ and $w_2 = 2$, analogous to the sketch in Fig.~\ref{fig-sketch}. The initial density is uniform, and the initial profile of the temperature is non-uniform, varying from 1 to 1.5, with a profile chosen to balance total pressure (plasma plus magnetic) to ensure the profile is in MHD equilibrium. The fluid pressure is provided fully by ions and is treated as adiabatic, while electrons are assumed cold at all times and carry all of the initial current. 

We initialize all simulations with a coherent perturbation in the magnetic field, for which the $z$ component of the magnetic vector potential $A_{1z}$ is
\begin{equation}
A_{1z}(x,y,z) = \frac{\tilde{B}_{1}}{4 \pi L_y}\left[1 + \cos\left(\frac{4\pi(y-L_y/4)}{L_y} \right) \right] \sin\left(\frac{2\pi x}{L_x}\right) f(z)
\label{eq-pert}
\end{equation}
for $y \geq 0$ and 0 for $y < 0$, where $\tilde{B}_{1} = 0.005$ is a constant and the envelope $f(z)$ has the form
\begin{equation}
f(z) = \frac{1}{2} \left[\tanh \left(\frac{z + w_{0pert}}{2}\right) - \tanh \left(\frac{z - w_{0pert}}{2}\right)\right], \label{eq-fofz}
\end{equation}
where $w_{0pert} = 15$ is the initial half-length of the coherent perturbation in the out-of-plane direction. The resulting magnetic perturbation ${\bf B}_1 = -{\bf \hat{z}} \times \nabla A_{1z}$ creates an x-line/o-line pair in the $xy$ plane for only the upper current sheet at $y = y_{cs} = L_y/4$, localized to $-w_{0pert} < z < w_{0pert}$. Figure~\ref{fig-initconds}(b) shows a cut in the $yz$ plane at $x=-L_x/2$ of the $y$-component of the coherent perturbation in the magnetic field. The value of $w_{0pert}$ is chosen to ensure the perturbation is localized exclusively in the region of half-thickness $w_1$ so that any reconnection observed in the region of half-thickness $w_2$ is due to spreading of reconnection and not due to the initial perturbation. We perturb only the upper current sheet to prolong the timescale for the interaction between the two current sheets resulting from flows in the $y$-direction and thus ensure the reconnection occurring in the upper sheet at later times is not caused by the lower current sheet.

Incoherent noise in the $x$ and $y$ components of the magnetic field at the $10^{-5}$ level is included to break symmetry, which prevents secondary magnetic islands from staying at the initial x-line location [{\it e.g.,} \cite{Shay04}]. A fourth-order diffusion term is included in all equations with coefficients $D_{4x} = D_{4y} = 1.6\times 10^{-5} \ d_{i0}^4 \Omega_{ci0}^{-1}$ in the $x$ and $y$ directions and a larger diffusion coefficient in the $z$ direction $D_{4z}=1.6\times 10^{-1}$ due to the larger grid scale. These values are varied in trial simulations to ensure they do not play any significant role in the numerics.

\section{Results}
\label{sec-results}

We begin by testing the spreading speed prediction in equations~(\ref{eq:caz}) and (\ref{nonuniform-speed-gradual-thick-thin}). First, we discuss how we find where reconnection is taking place in our 3D simulations and how we determine the speed at which the reconnection spreads. 

The strength of the normal magnetic field component $B_y$ near the reconnection region is an indicator of the presence of reconnection \cite{huba02, Jain17, tak, arencibia21}. The average magnitude of $B_y$ at the left and right downstream edges of the electron diffusion region is a proxy for the reconnection rate; we denote this quantity as $\tilde{B}_y(z,t)$, given by
\begin{equation}
 \tilde{B}_y(z,t) = \frac{\left|B_y\left(\tilde{x} +
   L,y_{cs},z,t\right)\right| + \left|B_y\left(\tilde{x} -
   L,y_{cs},z,t\right)\right|}{2}, \label{eq-byavgdef}
 \end{equation}
where $\tilde{x}$ is the $x$ location of the x-line in the plane specified by $z$ at time $t$ and $L \sim 2$ is the approximate half-length of the electron diffusion region. For further details see Sec. IV D in \citeA{arencibia21}. 

\begin{sidewaysfigure}
\center  
  \includegraphics{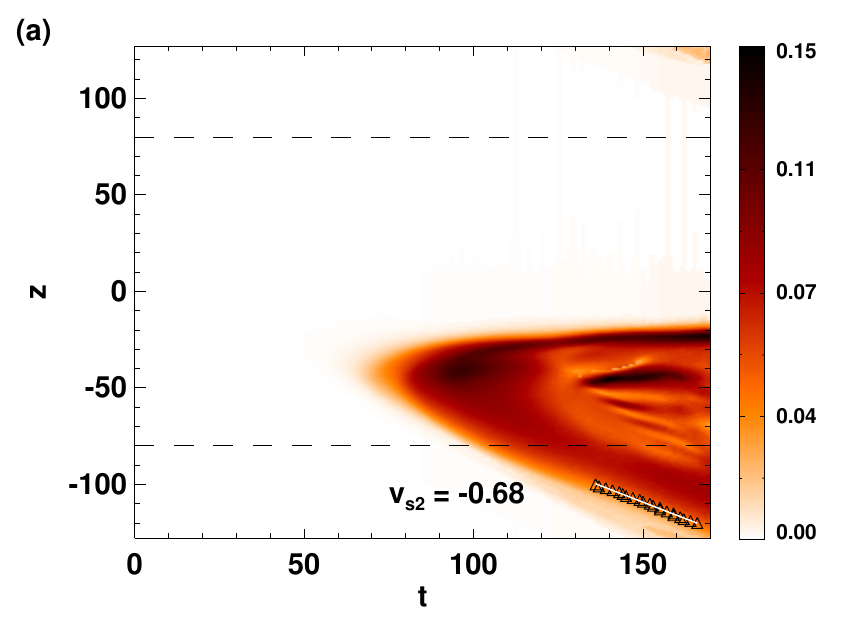} \includegraphics{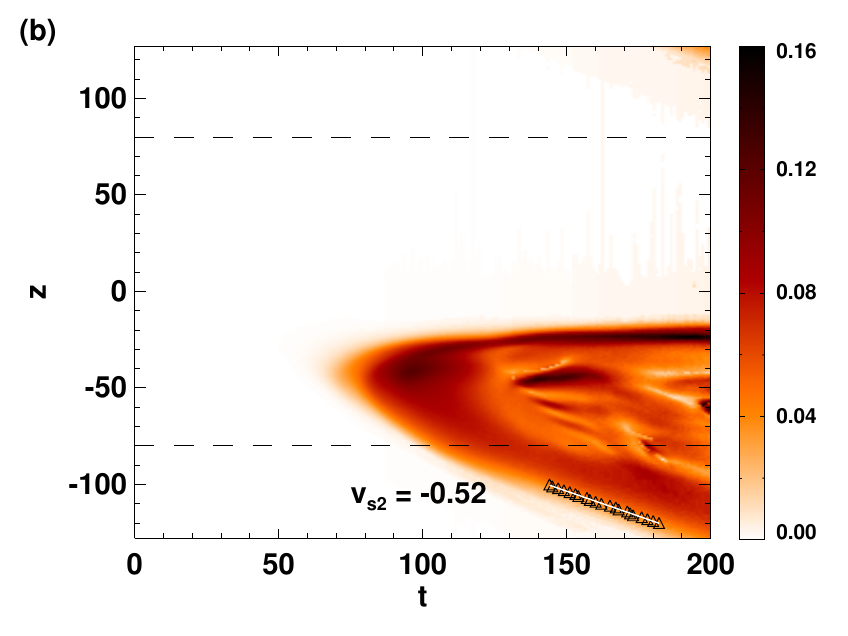}
  
  \includegraphics{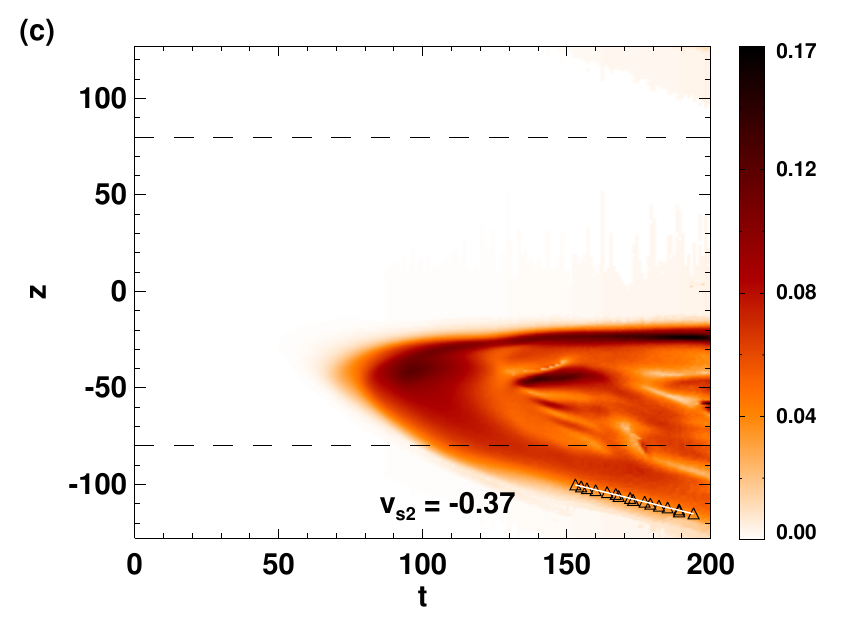} \includegraphics{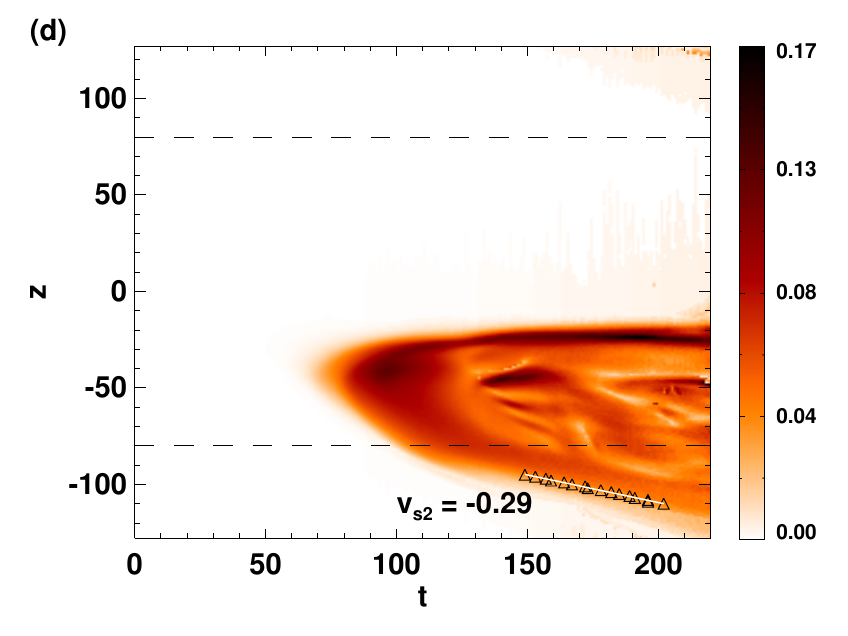}

  \caption{Average reconnected magnetic field $\tilde{B_y}(z,t)$, defined in equation~(\ref{eq-byavgdef}), as a function of the out-of-plane direction $z$ and time $t$, for simulations with initial non-uniform current sheet half-thicknesses $w_1 = 1$ for $-80<z<80$ and opening out to $w_2 = $ (a) 1.25, (b) 1.5 (c) 1.75, and (d) 2 elsewhere, where distances are in units of $d_{i0}$ and times are in $\Omega_{ci0}^{-1}$. The dashed lines separate the regions of different thicknesses, where $w_1$ is localized within $-80<z<80$ and $w_2$ elsewhere. Spreading in the thinner region of current sheet thickness $w_1$ occurs between $t = 80$ and 100. The spreading is slower in the region of thickness $w_2$. Black triangles denote when and where $\tilde{B_y} = 0.04$ for a chosen range in $z$ in the region of initial half-thickness $w_2$. The out-of-plane reconnection spreading speed $v_{s2}$ in the region with half-thickness $w_2$, listed for each simulation in units of $c_{A0}$, is the best fit slope of these points, shown as a white line. \label{fig-stacknu}}
\end{sidewaysfigure}

The average reconnected field $\tilde{B}_y(z,t)$ for the upper current sheet ($y = y_{cs} = L_y/4$) is shown in Fig.~\ref{fig-stacknu} as a stack plot as a function of time $t$ and out-of-plane coordinate $z$ over the whole domain for four 3D simulations with non-uniform thickness. Panels (a) through (d) have $w_2 = 1.25, 1.5, 1.75,$ and $2.0$ in the thicker part of the current sheet, respectively, and all four have $w_1 = 1$ in the thinner part. Each horizontal cut represents data from a fixed $xy$ plane as a function of time $t$, while each vertical cut represents the spatial extent of $\tilde{B}_y$ in the $z$ direction of the reconnecting region at a fixed time. The triangular shape of $\tilde{B}_y(z,t)$ seen in Fig.~\ref{fig-stacknu} is a characteristic of reconnection that is spreading uni-directionally \cite{Shay03,Shepherd12}, as the extent of the reconnection region increases in time. We see $\tilde{B}_y$ increase in time from 0 to an asymptotic value of $\approx 0.1$ when reconnection reaches a quasi-steady state in the current sheet region with local half-thickness $w_1$ before spreading in the $-z$ direction into the region with local half-thickness $w_2$.

We define the onset of fast reconnection at a given $xy$ plane to be when $\tilde{B}_y$ exceeds 0.04, after which reconnection proceeds to a quasi-steady state \cite{arencibia21}. Onset times for individual $xy$ planes are plotted in Fig.~\ref{fig-stacknu} as black triangles for a chosen range of $z$ values in the region of initial half-thickness $w_2$. The spreading speeds $v_{s2}$ in the region of uniform half-thickness $w_2$ are simply the slope of the collection of points denoting the onset time. We determine this slope using a least squares fit and the slopes are shown as the white lines in each panel of Fig.~\ref{fig-stacknu}. The spreading speed $v_{s1}$ in the region of half-thickness $w_1 = 1$ for all four simulations in Fig.~\ref{fig-stacknu} are $\simeq 1.0$ (not shown), which is consistent with equation~(\ref{eq:vsuniform6}) as expected \cite{huba02,Shay03}. In all four cases, there is a break in the spreading speed where the reconnection reaches the region of larger half-thickness $w_2$ and all show spreading speeds well below $c_{A0}d_{i0}/w_2$, which would be the expected spreading speed if the uniform current sheet spreading speed theory \cite{huba02,Shay03} was valid for current sheets of non-uniform thickness. 

\begin{table}
\centering
\caption{Results for 3D two-fluid simulations in this study. The first column gives ordered pairs $(w_1,w_2)$ for current sheets that vary in half-thickness along the out-of-plane direction from a value of $w_1$ to $w_2$ in units of $d_{i0}$. $v_{s2}$ is the reconnection spreading speed in the region with half-thickness $w_2$. The second column gives the theoretical predictions from   Sec.~(\ref{sec-theory}), and the third column gives the values measured from the simulations. The deviation from the theory is shown as a percentage in the fourth column.}
\label{table1}
\begin{tabular}{lllll}
$(w_1,w_2)$ & \ Predicted $v_{s2}$ & Measured $v_{s2}$ & Deviation  &   \\ \hline
(1.0,1.0)       & 1.00                     & 0.97              & -3.1\%  &   \\
(2.0,2.0)       & 0.50                     & 0.51              & 2.0\%  &   \\ \hline
(1.9,2.0)       & 0.48                     & 0.41              & -15.9\% &   \\
(1.75,2.0)      & 0.44                     & 0.38              & -15.1\% &   \\
(1.5,2.0)       & 0.38                     & 0.31              & -21.0\% &   \\
(1.0,2.0)       & 0.25                     & 0.29              & 13.8\% &   \\ (0.75,2.0)      & 0.19                     & 0.26              & -27.9\% &   \\
\hline
(1.0,1.25)      & 0.64                     & 0.68              & 5.9\%  &   \\
(1.0,1.5)       & 0.44                     & 0.52              & 14.5\% &   \\
(1.0,1.75)      & 0.33                     & 0.37              & 11.7\% &   \\ \hline
(2.25,2.0)      & 0.50                        & 0.46              & -8.0\%  &   \\
(2.5,2.0)       & 0.50                        & 0.51              & 2.0\%  &   \\
(2.0,1.5)       & 0.67                        & 0.73              & 9.5\%  
\end{tabular}
\end{table}
Stack plots analogous to those in Fig.~\ref{fig-stacknu} are generated and spreading speeds are obtained using the same method for all the simulations in this study (not shown). Table~\ref{table1} gathers the results for all simulations in this study in the first column, labeled as ordered pairs $(w_1,w_2)$ according to their respective current sheet half-thicknesses. We include the spreading speed prediction from equations~(\ref{eq:caz}) and (\ref{nonuniform-speed-gradual-thick-thin}) in the region of half-thickness $w_2$ in the second column for simulations with $w_1 < w_2$ and $w_1 > w_2$, respectively. The third column is the calculated spreading speed magnitude $v_{s2}$ from the simulations and the fourth column is the deviation from the theoretical prediction shown as a percentage. 

\begin{figure}
\includegraphics[width=\textwidth]{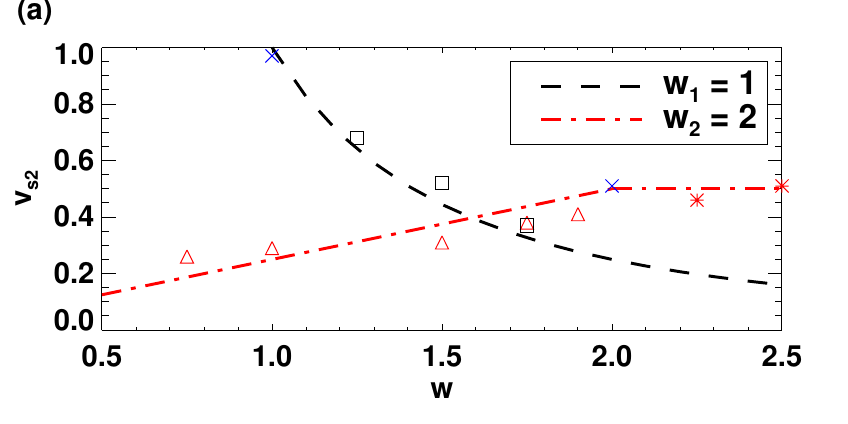}
\includegraphics[width=\textwidth]{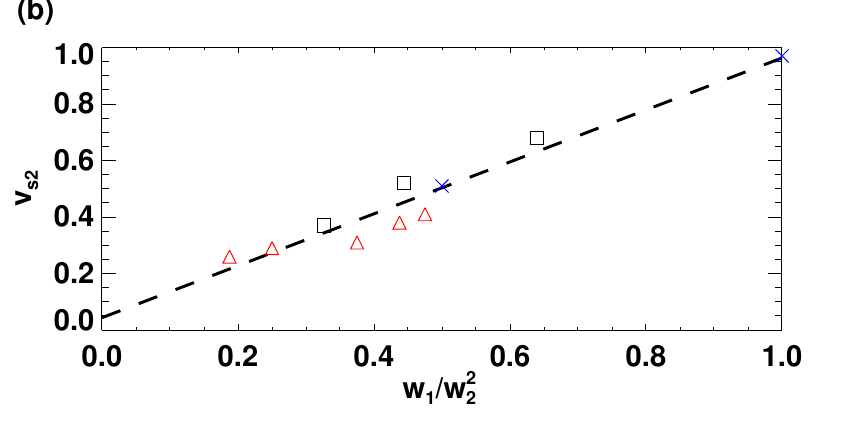}
  \caption{Comparison of simulation results and theory for the reconnection spreading speed $v_{s2}$ in units of $c_{A0}$ in the region where the initial current sheet half-thickness is $w_2$ in units of $d_{i0}$ for anti-parallel reconnection simulations with current sheets that vary in the out-of-plane direction from a half-thickness $w_1$ to $w_2$.  (a) $v_{s2}$ as a function of the current sheet half-thickness $w$, which represents either independent variables $w_1$ or $w_2$, depending on which is the independent variable for the given set of simulations.  Two uniform half-thickness simulations $(w_1, w_2) = (1, 1)$ and (2, 2) are shown as blue crosses. Three simulations with $w_1$ = 1 and $w_2$ = 1.25, 1.5, 1.75 are shown as black squares and the independent variable is $w = w_2$. Five simulations with $w_1$ = 0.75, 1.0, 1.25.1.5, 1.75, 1.9 and $w_2$ = 2 are shown as red triangles and two simulations with $w_1$ = 2.25, 2.5 and $w_2$ = 2 are shown as red asterisks, where the independent variable is $w = w_1$. The dashed black line represents the theoretical prediction from equation~(\ref{eq:caz}) for simulations with $w_1 = 1$ fixed with $w=w_2$ as the independent variable.  The red dash-dot piecewise-curve represents the prediction from equation~(\ref{eq:caz}) with $w_2 = 2$ fixed with $w=w_1$ as the independent variable for $w_1 < w_2$, and equation~(\ref{nonuniform-speed-gradual-thick-thin}) for $w_1 > w_2$. (b) $v_{s2}$ as a function of $w_1/w_2^2$ for simulations with $w_1 \leq w_2$.  The dashed black line gives the theoretical prediction from equation~(\ref{eq:caz}).
  \label{fig-scalingnu}}
\end{figure}
We gather the spreading speeds from our simulations in Fig.~\ref{fig-scalingnu}. Panel (a) shows $v_{s2}$ as a function of the current sheet half-thickness $w$, which represents either independent variables $w_1$ or $w_2$, depending on which is the independent variable for the given set of simulations. The two uniform half-thickness simulations $(w_1,w_2) = (1,1)$ and $(2,2)$ are shown as blue crosses. The three simulations with $w_1=1$ and $w_2=1.25,1.5,1.75$ are shown as black squares and the independent variable is $w=w_2$. The five simulations with $w_1=0.75,1.0,1.25.1.5,1.75,1.9$ and $w_2=2$ are shown as red triangles and the two simulations with $w_1=2.25,2.5$ and $w_2=2$ are shown as red asterisks, where the independent variable is $w=w_1$. The dashed black line represents the theoretical prediction from equation~(\ref{eq:caz}) for simulations with $w_1 = 1$ fixed with $w=w_2$ as the independent variable.  The red dash-dot piecewise-curve represents the prediction from equation~(\ref{eq:caz}) with $w_2 = 2$ fixed with $w=w_1$ as the independent variable for $w_1 < w_2$, and equation~(\ref{nonuniform-speed-gradual-thick-thin}) for $w_1 > w_2$. The simulation results are in excellent agreement with the theory. The $(w_1,w_2)=(2,1.5)$ simulation is not expected to lie on either of the two curves and thus is not shown. 

To test the agreement more quantitatively, Fig.~\ref{fig-scalingnu}(b) shows spreading speeds $v_{s2}$ for all simulations with $w_1 \leq w_2$ as a function of $w_1/w_2^2$, the predicted dependence from equation~(\ref{eq:caz}). We calculate a linear least squares fit of these points and show the fit as a dashed line with a functional form $v_{s2} = (0.919\pm0.082)w_1/w_2^2 + (0.044\pm0.041)$, showing excellent agreement with equation~(\ref{eq:caz}). Simulations with $w_1 > w_2$ are not included in the fit as they are predicted to satisfy a different scaling. We conclude the theory of spreading speeds in a current sheet varying in thickness from $w_1$ to $w_2$ are consistent with the predictions in Sec.~\ref{sec-theory}.

We use the same simulations to test our prediction for spreading in current sheets with a thickness that varies continuously in the out-of-plane direction. To compare with equation~(\ref{eq-tau}), we estimate the spreading timescale in the region where the current sheet thickness changes in the simulation with $(w_1,w_2)=(1,2)$. The stack plot for the simulation in Fig.~\ref{fig-stacknu}(d) shows that reconnection spreads across the region $-84<z<-76$ approximately over the time range $100<t<120$, so $\tau \simeq 20$.
From equation~(\ref{eq-wofz}), the thickness varies approximately linearly across the transition region $-84<z<-76$, so we use $\alpha=1$ and $\Delta z=2w_z=8$. Using equation~(\ref{eq-tau}), the spreading time across the region where the current sheet thickness changes is predicted to be $\tau\approx19$. This is in good agreement with the simulation results.  To further test the theory, two additional simulations with $(w_1,w_2)=(1,2)$ are performed using $w_z=8$ and $w_z=12$ for the gradient length scale in equation~(\ref{eq-wofz}), doubling and tripling $\Delta z$. The spreading timescales in the higher $w_z$ simulations increase approximately by factors of 2 and 3, respectively (not shown). This is in agreement with the predicted scaling with $\Delta z$ in equation~(\ref{eq-tau}) assuming the same linear profile with $\alpha=1$.  These results suggest that the theory for the spreading speed in current sheets with a gradually varying thickness is valid.

\begin{figure} 
\center \includegraphics[width=\textwidth]{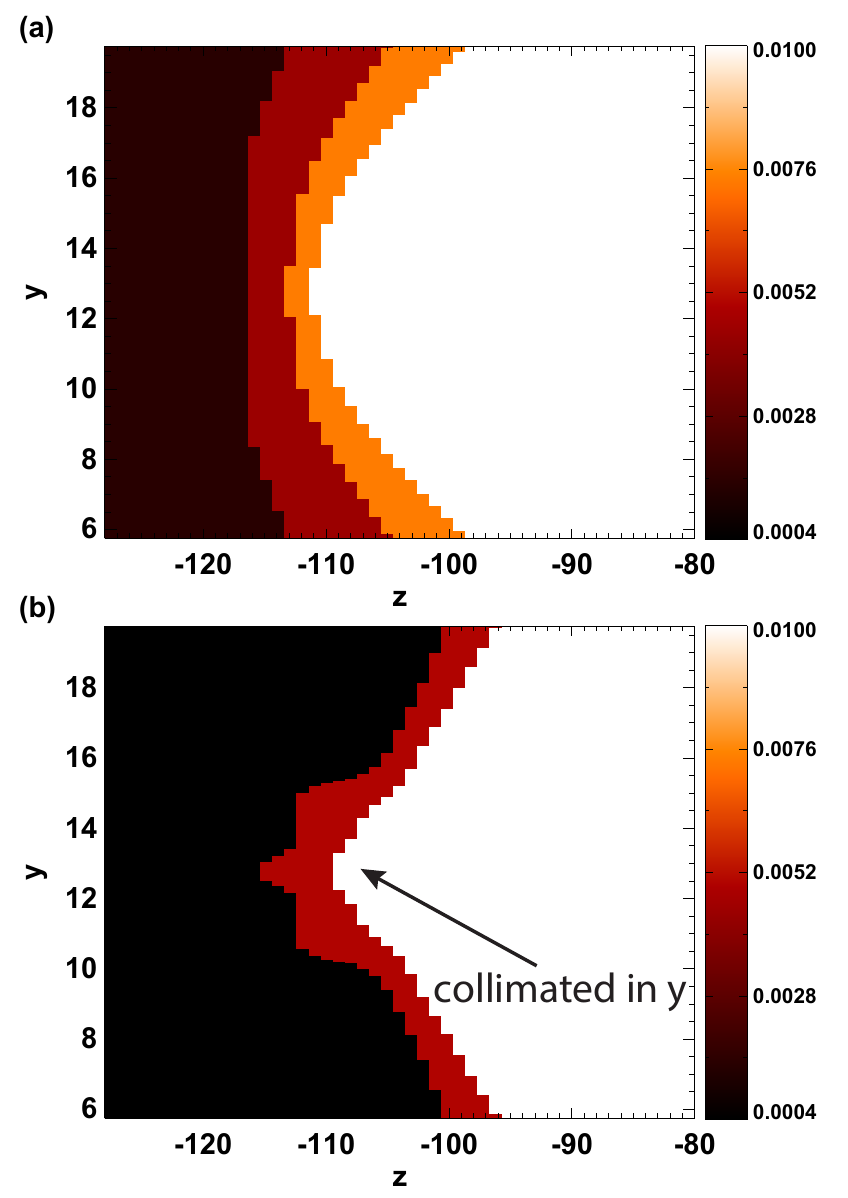}
  \caption{Reconnected magnetic field component $B_y$ in cuts in the $yz$ plane at $x = \tilde{x} - L$ for simulations with (a) a current sheet with uniform initial half-thickness of $w_0 = 2$ at $t=280$ and (b) a current sheet with non-uniform initial thickness   with $w_2 = 2$ for $-128<z<-80$ and $w_1 = 1$ for $-80 < z < 80$ at $t = 170$.  $B_y$ is collimated in the $y$ direction when spreading from a thinner into a thicker current sheet. }
  \label{fig-byyz}
\end{figure}
Finally, we confirm the assumption that the reconnected magnetic field $B_y$ is collimated for spreading from a thinner to a thicker current sheet. We do this by a visual inspection of $B_y$ in the thick part of a non-uniform current sheet and compare it with a simulation with a uniform and equal local thickness. Figure~\ref{fig-byyz} shows cuts of $B_y$ in the $yz$ plane at $x = \tilde{x} - L$ where $L = 2$ is the approximate half-length of the electron diffusion region, taken at representative times when $B_y$ approximately reaches the $z = -115$ plane with an amplitude comparable to the initial perturbation $\sim 0.004$ for (a) a simulation with uniform thickness $w_0 = 2$ at $t = 280$ and (b) the simulation with $w_1 = 1$ and $w_2 = 2$ at $t = 170$. The color bar is saturated at a relatively weak value of $B_y = 0.01$ to resolve the very weak magnetic fields spreading in the out-of-plane direction (right to left). Focusing on the saturated (white) colors, panel (b) shows that $B_y$ is significantly collimated in the $y$ direction around the current sheet center $y = 12.8$ when compared to the uniform case in panel (a), hence initially perturbing only a smaller embedded portion of the magnetic field, illustrating the assumption of the theory. A similar comparison was carried out between the simulation with uniform thickness $w_0 = 2$ at $t = 280$ and the simulation with $w_1 = 1.5$ and $w_2 = 2$ at $t = 255$ (not shown). For this comparison with more similar thicknesses $w_1$ and $w_2$, no appreciable difference in the thickness of the convected $B_y$ could be seen in the two simulations. We attribute this to the spatial scale of the collimation being too similar for differences to be resolved. Uncertainties in the time slices used for the comparison also make it challenging to discern the collimation of $B_y$ for this comparison. In summary, a simulation with sufficient scale separation shows clear evidence that the reconnected $B_y$ is collimated as it convects.

\section{Applications}
\label{sec-app}

\subsection{Reconnection Spreading in the Near-Earth Magnetotail}
\label{sec-tailspreading}

The central plasma sheet in the near-Earth magnetotail is known to vary in thickness continuously in the dawn-dusk direction $Y_{GSM}$ in the Geocentric Solar Magnetospheric (GSM) coordinate system, equivalent to the $z$ coordinate in Sec.~\ref{sec-theory}. It has a minimum thickness at midnight magnetic local time and maxima at the flanks near the nightside magnetopause [see Fig.~3 in \citeA{voigt84} and Fig.~7 in \citeA{tsyganenko98}]. Thin current sheets are more commonly found from midnight to dusk \cite{Rogers22}.  The anti-parallel field configuration and low collisionality makes it an ideal system to apply the theory. 

First, we compute the timescale for magnetic reconnection to spread during a substorm expansion event in the near-Earth magnetotail. For pre-substorm initial conditions, assuming a fully ionized hydrogen plasma, a reconnecting magnetic field with an asymptotic value $B_x \approx 20$ nT \cite{miyashita20} and a magnetosphere density at the plasma sheet boundary layer of $n \approx 0.1$~cm$^{-3}$ \cite{baumjohann90}, we estimate the ion inertial scale is $d_i \approx 720$~km and the Alfv\'en speed is $c_A\approx1400$~km s$^{-1}$. We take $w_1 \approx 0.1-0.4 R_E \approx 0.89-3.5 \ d_i$ as the minimum cross-tail half-thickness at midnight and $w_2 \approx 1 R_E \approx 8.9 \ d_i$ as the maxima at the flanks, and $\Delta z\sim15R_E\approx133d_i$ as the approximate half-length of the cross-tail current sheet along the dawn-dusk direction \cite{fairfield80,sergeev90}. Since $\Delta z$ greatly exceeds the ion inertial scale, the assumption that the current sheet only gradually becomes thicker is valid.  We assume reconnection begins with a finite x-line with its dawnward edge situated at midnight ($Y_{GSM} = 0$), such that reconnection spreads dawnwards in the direction of electron motion \cite{Nagai11,Nagai13} until reaching $Y_{GSM} = -\Delta z = -15R_E$. Assuming a parabolic cross-tail current sheet, we use $\alpha=2$ in equation~(\ref{eq-tailprofile}). Using equation~(\ref{eq-tau}), this gives spreading timescales in the range $\tau \approx2.5-23.6$ minutes, where the range depends on the value for $w_1$. For comparison, if the cross-tail current sheet were uniform with a typical midnight half-thickness $w_1 \approx 0.1-0.4 \ R_E \approx 0.89-3.5 \ d_i$, equation~(\ref{eq:vsuniform6}) implies the timescale for spreading would be in the range of $\tau \approx 1-4$ minutes, comparable to the Alfv\'en crossing time $\approx 1$ minute. Thus the theory provides a mechanism for reconnection spreading along the cross-tail current sheet on timescales longer than both what Alfv\'en and current carrier speeds suggest.

Observations also suggest that reconnection in the near-Earth magnetotail may begin with an x-line with its dawnward edge at $Y_{GSM}>0$ \cite{Shay03,Nakamura04,Nagai13}. This suggests reconnection may first spread from a thicker part of the current sheet into the thinner part at midnight before continuing to spread dawnwards towards a thicker part of the current sheet. In this scenario, the timescale for spreading in the $Y_{GSM}>0$ region would be calculated with equation~(\ref{eq-tau2}) for spreading along the region of decreasing current sheet thickness.

There is also observational evidence that reconnection may not spread across the entire dawn-dusk direction, instead stopping when the x-line is $\sim8R_E$ in length \cite{Nagai13}. Constraining $\Delta z$ to empirical values in equations~(\ref{eq-tau}) and (\ref{eq-tau2}) may give more accurate predictions. We point out that the structure of the cross-tail current sheet may also be more complex and bend away from the dawn-dusk direction asymmetrically near the flanks due to seasonal and diurnal oscillations of Earth's dipole tilt angle \cite{tsyganenko98}. This effect is not captured in our model current sheet, but it is reasonable to expect that if the radius of curvature of the plasma sheet is much larger than the ion inertial scale that it would introduce only small corrections to the present results.

\subsection{Reconnection Spreading in Two-Ribbon Solar Flares}
\label{sec-coronalspreading}

Another scenario where the theory may be applicable is in the spreading or ``zipper'' motion of the ribbons in two-ribbon flares, which is thought to result from out-of-plane spreading of magnetic reconnection in the solar corona [see \citeA{Qiu10,Tian15,Qiu17} and references therein]. \citeA{Qiu17} analyzed six two-ribbon flare events that show ribbon elongation/spreading occurs at speeds typically slower than the coronal Alfv\'en speed by as much as an order of magnitude. One previously known mechanism that could explain a sub-Alfv\'enic reconnection spreading speed is that the current sheet could have uniform thickness but be thicker than ion inertial scales \cite{Shay03,arencibia21}. This may be a potential explanation for unidirectional spreading of ribbons with a uniform speed in flare events with a weak guide field, such as in Fig.~5 in \citeA{Qiu17}. 

The results of the present study provide another mechanism for spreading speeds below the Alfv\'en speed. An observational signature of this scenario is a reconnection spreading speed that slows with distance. Additionally, if the minimum and maximum half-thicknesses $w_1$ and $w_2$ are both larger than the ion inertial scale, the spreading speed is predicted to be both sub-Alfv\'enic and below the local current carrier speed at any location in the current sheet. This is qualitatively similar to the behavior of observed ribbon elongation speeds in Fig.~2 in \citeA{Qiu17}  and Fig.~9 in \citeA{naus22}, both showing ribbon elongation speeds varying along the direction of spreading, although we note the former is for an event in which the flare ribbons spread in the direction opposite to that of the inferred current carriers. This signature may potentially be useful for inferring the structure of a reconnecting coronal current sheet as has been alluded to in \citeA{naus22}, even though the thicknesses in question are far below currently resolvable scales in the corona ($d_i\sim10$ m). 

\section{Conclusions}
\label{sec-conc}

We develop a scaling theory of collisionless magnetic reconnection spreading for anti-parallel reconnection with current sheet thicknesses that vary in the out-of-plane direction. Existing theories only apply to current sheets of uniform thickness, predicting that anti-parallel collisionless reconnection spreads at the speed of the local current carriers in the sheet, $v_s = c_A(d_i/w_1)$ for a current sheet of uniform half-thickness $w_1$, where $c_A$ is the Alfv\'en speed based on the reconnecting field and $d_i$ is the ion inertial length. For non-uniform thickness sheets for which reconnection initiates where the half-thickness is $w_1$ that spreads into a thicker current sheet of half-thickness $w(z)$, we predict that the spreading speed is reduced to $v_s = c_Aw_1d_i/[w(z)]^2$, {\it i.e.,} by a factor of $w_1/w(z)$, due to a reduction in the initial effective reconnecting magnetic field strength \cite{Shay04}. Therefore, there is a memory effect from the region from which reconnection starts. Importantly, our result provides a mechanism for reconnection spreading slower than the Alfv\'en and current carrier speeds, which has been inferred from observations in both the solar and magnetospheric settings. For spreading from a thicker to thinner current sheet, the spreading speed is the speed of the current carriers, $v_s = c_A d_i/w(z)$, so there is no memory effect from the region that reconnection begins. We perform a calculation of the time-scale of reconnection spreading in a current sheet with a known profile for $w(z)$. We confirm our predictions with 3D two-fluid numerical simulations. 

We apply our results to physical systems where the thickness of reconnecting current sheets is known or expected to be initially non-uniform in the out-of-plane direction. In Earth's magnetotail, where the thickness of the near-Earth cross-tail current sheet increases continuously from midnight out to the flank magnetopause, using a model magnetotail shape at active times provides a prediction of the time scale for the spreading. Such an analysis could also be employed for quiet time events, but this was not carried out here. Both predictions should be able to be compared with direct or remote observations, which would be an important step for future work.  In two-ribbon solar flares, our result may potentially explain why the ribbons in events with nearly anti-parallel reconnecting fields may spread at sub-Alfv\'enic speeds. Moreover, we provide an observational signature for spreading in a current sheet with a varying thickness, {\it i.e.,} that the speeds change in time during the spreading process. The inferred current sheet thicknesses remain far below current observational capacities, so other approaches will be necessary to confirm or refute the model in solar flares. 

There are a number of other avenues for future studies. Our simulations assume the asymptotic reconnecting magnetic field strength is the same everywhere along the current sheet, but this need not be the case.  We expect that the results here would carry over with $B_0$ replaced by $B_x(z)$ in such a scenario, but future work would be required to test this hypothesis. Simulations in a 3D box geometry may leave out important geometrical effects from realistic systems, including curvature of the magnetic fields and density structure in the solar corona, as well as curvature of the near-Earth magnetotail current sheet during seasonal and diurnal oscillations of the Earth's dipole tilt angle and the normal $B_{z,GSM}$ present in the near-Earth magnetotail.  Our study does not include an out-of-plane (guide) magnetic field, which may be relevant in solar flare ribbon spreading events and for the dayside magnetopause and the solar wind. An extension of our results to asymmetric reconnection may also be useful for the study of reconnection spreading at the dayside magnetopause, where it has been reported that the spreading speed of reconnection is sub-Alfv\'enic \cite{Zou18}. Generalizing the result to asymmetric reconnection would be necessary to test whether the mechanism discussed here explains the decrease in spreading speed at the dayside magnetopause seen in \citeA{walsh18}, where the current sheet is thinnest near the nose and gets thicker as one goes downtail. The present simulations employ cold electrons within the two-fluid model, so drift waves are absent.  In a realistic system, drift waves are expected to potentially be excited where there is a change in the current sheet thickness in the out-of-plane direction.  It may be interesting to study reconnection spreading in systems in which the current sheet thickness changes on kinetic scales to see if drift waves play a role and to determine if the equilibrium current prevents spreading. The effect might be expected to be small if the current sheet thickness changes over length scales larger than the electron inertial scale, but studying whether drift waves impact the spreading speed should be the subject of future extended-MHD or kinetic modeling. It would also be interesting to more rigorously describe the effects of embedding in the theory in Sec.~\ref{subsec-thintothick}.

\section{Open Research}

The simulation study was carried out using the two-fluid code F3D \cite{Shay04}. The model parameters used are detailed in Sec.~\ref{sec-simulation}. Data analysis was carried out, and all simulation figures were generated, with IDL 8.2. Processed simulation data supporting the results and used to generate all simulation figures is publicly available \cite{Arencibia22data,Arencibia22data-add}.







\acknowledgments
We acknowledge helpful conversations with Dana Longcope, Toshi Nishimura, Eric Priest, Kathy Reeves, and Luke Shepherd. We thank Mahmud Hasan Barbhuiya for assistance with annotating Figure~\ref{fig-byyz}. Support from NSF Grants AGS-1460037 (PAC), AGS 1602769 (PAC), AST-1839084 (JQ), AGS-2024198 (MAS), OIA-1655280 (HL), DOE Grant DE-SC0020294 (PAC), NASA Grants 80NSSC19M0146 (PAC), NNX16AG76G (PAC), 80NSSC18K1379 (SMP), 80NSSC20K1813 (MAS), SUB000313/80GSFC19C0027 (HL), SV4-84017 (HL), and 80NSSC21K0003 (HL), and contract 499935Q (SMP) is gratefully acknowledged.  Computational resources supporting this work were provided by the NASA High-End Computing (HEC) Program through the NASA Advanced Supercomputing (NAS) Division at Ames Research Center and by the National Energy Research Scientific Computing Center (NERSC), a DOE Office of Science User Facility supported by the Office of Science of the U.S.~Department of Energy under Contract No.~DE-AC02-05CH11231.


%
%



\bibliography{gcrbib}

%
%
%
%
%

\end{document}